\documentclass[runningheads]{llncs}

\usepackage{float}
\usepackage{graphicx} 
\usepackage{amsmath}
\usepackage{amssymb}
\usepackage{bbm}
\usepackage{mathtools}
\usepackage{makecell}
\usepackage{authblk}  
\usepackage[textsize=tiny]{todonotes}
\usepackage[hypertexnames=false,hyperfootnotes=false]{hyperref}
\usepackage{cleveref}
\usepackage{url}
\usepackage{ifthen}
\pdfoutput=1

\usepackage{bbm}
\usepackage{tikz}
\usepackage{amsmath}
\usepackage{physics}
\usepackage{graphicx}
\usepackage{setspace}
\usepackage{mathtools}

\usepackage[font={small}]{caption}

\makeatletter
\def\th@plain{%
  \thm@notefont{}
  \itshape 
}
\def\th@definition{%
  \thm@notefont{}
  \normalfont 
}
\makeatother

\usepackage{sectsty}

\newtheorem{assumption}[theorem]{\text{Assumption}}

\newcommand{\E}{\mathbb{E}}

\newcommand{\R}{\mathbb{R}}
\newcommand{\1}{\mathbbm{1}}

\renewcommand{\d}{\partial}

\newcommand\blfootnote[1]{%
  \begingroup
  \renewcommand\thefootnote{}\footnote{#1}%
  \addtocounter{footnote}{-1}%
  \endgroup
}

\numberwithin{theorem}{section}
\numberwithin{proposition}{section}
\numberwithin{corollary}{section}

\title{Information Structures in Stablecoin Markets}
\titlerunning{Redemptions and Runs}

  \author{
    Brian Z. Zhu \inst{1} 
  }
  \institute{Columbia University}

  \authorrunning{Zhu}

\begin{document}

\maketitle

\begin{abstract}
Stablecoins have historically depegged due from par to large sales, possibly of speculative nature, or poor reserve asset quality. Using a global game which addresses both concerns, we show that the selling pressure on stablecoin holders increases in the presence of a large redemption. While precise public knowledge reduces (increases) the probability of a run when fundamentals are strong (weak), interestingly, more precise private signals increase (reduce) the probability of a run when fundamentals are strong (weak), potentially explaining the stability of opaque stablecoins. The total run probability can be decomposed into components representing risks from large sales and poor collateral. By analyzing how these risk components vary with respect to information uncertainty and fundamentals, we can split the fundamental space into regions based on the type of risk a stablecoin issuer is more prone to. We suggest testable implications and connect our model's implications to real-world applications, including depegging events, stablecoin regulations, and the no-questions-asked property of money.
\blfootnote{We are grateful for the support of Pantera Capital's Catalyze Research Fellowship.}
\end{abstract}

\section{Introduction}\label{sec:intro}

Stablecoins are blockchain-based cryptographic tokens whose price is supposed to be at par with respect to a fiat currency, most commonly the US dollar. To achieve this peg, stablecoin issuers claim to back each token with at least one dollar worth of reserve assets. The decentralization and lack of regulation around stablecoins means that issuers can choose from a large range of assets to hold as reserves and control the transparency or public perception of their reserves to a great extent. The range of assets has included US Treasuries, bank deposits, money market instruments, cryptocurrencies, and abstract ``assets'' like pricing algorithms. Some issuers publicly disclose their reserves, hiring custodians to monitor and maintain those reserves. Other issuers do not disclose verifiable information regarding their reserves, intentionally or not, leading to uncertainty among investors.

The choice of reserve assets and degree of uncertainty about them is related to types of risk that a stablecoin issuer faces. We provide a few examples. In March 2023, Circle, who issues USD Coin (USDC), disclosed that \$3.3 billion of its reserves were held at Silicon Valley Bank (SVB) admist its collapse. This led to an immediate depeg, with USDC's peg being restored only when the US government announced a bailout of SVB. In this case, a transparent stablecoin whose reserves are known fell vulnerable to credit events occurring to the issuer's reserve assets.\footnote{https://fortune.com/crypto/2024/01/31/tether-loans-attestation-zero-paolo-ardoino-fourth-quarter/}\textsuperscript{,}\footnote{https://www.coindesk.com/layer2/2022/05/24/tether-transparency-is-needed-following-terras-ust-collapse-analyst/} Tether (USDT), issued by Tether Limited, makes less frequent reserve disclosures and faces scrutiny over its opaqueness regarding reserves. USDT has historically faced speculative attacks (for example, in October 2018 and May 2022) but has recovered from them each time, with no large-scale collapse occurring.\footnote{https://www.nasdaq.com/articles/op-ed:-anatomy-of-the-tether-attack:-are-stablecoins-vulnerable-2018-11-14}\textsuperscript{,}\footnote{https://thenextrecession.wordpress.com/2022/05/15/crypto-untethered/} TerraUSD (UST), issued by Terraform Labs, uses a supply-adjusting algorithm to maintain its price stability, resulting in an endogenously-backed stablecoin where its value depends on the stability of the token itself. However, selling pressure in May 2022 started a vicious cycle that ultimately led to the collapse of UST. There is debate on whether this selling pressure was due to a speculative attack.\footnote{https://www.richmondfed.org/publications/research/economic\_brief/2022/eb\_22-24}

Motivated by these observations, this paper studies the relationship between information uncertainty and the risk from large sales (possibly caused by speculative attacks) or poor collateral (possibly caused by credit events). As of November 2025, stablecoins have a market capitalization of around \$300 billion.\footnote{https://www.coingecko.com/en/categories/stablecoins} As they form the backbone of many DeFi services (such as automated market makers or lending protocols) and potentially have spillover effects on traditional financial services, it is crucial to understand this relationship. How information uncertainty affects run risk could guide policymaking and regulations on the disclosure and monitoring of reserves. Indeed the implications of our model can be used to analyze the effect of recent stablecoin legislation, such as the GENIUS Act (U.S.) and MiCA regulations (E.U.), that have increased requirements for stablecoin issuers.

To capture both types of risk, we extend a streamlined model of \cite{AhmedEtAl} to include an large agent that sells at random. To motivate the seller's randomness, we note that large sales can occur for a variety of reasons, including speculation and the liquidity needs of a large agent, yet small agents will perceive a degree of induced selling pressure regardless. This can also be viewed as repurposing the model of \cite{CorsettiEtAl} for stablecoins and relaxing it by making the large agent non-strategic. This relaxation allows us to establish uniqueness results and to derive novel comparative statics results with respect to the uncertainty in information of a stablecoin's reserve asset quality. 

We first show that a unique equilibrium in switching strategies exists both when a large sale occurs and does not occur; in this equilibrium, the selling pressure and conditional run probability increase in the presence of a large sale. We then decompose the run probability into two components: the conditional run probability given a large sale occurring and not occurring. The former component can be interpreted as the risk from by a large sale, and the latter as the risk from poor collateral. These conditional run probabilities decrease (increase) as public knowledge is more precise when fundamentals are sufficiently strong (weak). Intriguingly, as private signals become more precise, conditional run probabilities increase (decrease) for sufficiently strong (weak) fundamentals. The analysis with respect to private signals yields a novel result which suggests another potential source of a stablecoin's stability, i.e. a higher degree of heterogeneity in how investors idiosyncratically assess a stablecoin's reserves. Tokens that have large information uncertainty in this sense may thus benefit from it.

Isolating the different types of risk facing a stablecoin also allows us to partition the fundamental space into regions where the conditional run probabilities have differing behaviors with respect to both types of information uncertainty. This classification could be of possible use in assessing what types of risk a stablecoin is prone to given their quality of fundamentals and level of transparency about reserves. The implications generated by our model can be used to explain the events described earlier, and to provide more general insights about the stability mechanisms and regulatory challenges facing stablecoins. 

\paragraph{Institutional Details.} After the advent of the first cryptocurrencies in the mid-2010s, stablecoins were created to meet the demand for an on-chain asset with relative price stability. Since then, stablecoins have grown in market capitalization, peaking in early 2022 and hovering around \$300 billion as of January 2026. The uses of stablecoins are numerous, including cross-border funds transfers, stores of values for lending protocols, collateral for trading cryptocurrencies, and a medium of exchange in automated market makers.

The market microstructure of stablecoins can be split into the \textit{primary market} and \textit{secondary market}. In the primary market, the issuer interacts with arbitrageurs and reserve asset markets. The issuer holds a collection of reserve assets that ``back up'' the amount of stablecoins in circulation, and exchanges a minted token to arbitrageurs for a par value of the fiat currency that the stablecoin is pegged to. Arbitrageurs can redeem tokens at par, assuming that the issuer is solvent (i.e.\! can liquidate enough reserves to fulfill all redemption requests at a given time). Notably, the issuer has sole discretion over who can be an arbitrageur, which usually involves a vetting process. In the secondary market, the arbitrageurs trade with buyers and sellers at market prices on exchanges.

In July 2025, the United States government passed the GENIUS Act, establishing a comprehensive framework for payment stablecoins. Notably, to be a permitted stablecoin issuer, an entity must maintain full reserve backing of the token with money market assets (e.g.\ dollars, Treasures, ``similarly liquid assets''), publicly disclose their token redemption policy, and publish monthly audits or reports of their reserve composition. The legislation does not provide deposit insurance or Fed backstops for stablecoins, instead placing responsibility on the issuer to manage reserves and disclose information.\footnote{See https://www.congress.gov/bill/119th-congress/senate-bill/394  for the full text.}


\paragraph{Related Literature.} We contribute to the literature on stablecoin runs. On the game-theoretic side are the works of \cite{GortonEtAl}, \cite{MaEtAl}, \cite{Bertsch}, and \cite{AhmedEtAl}, which all use global games. \cite{GortonEtAl} shows that despite the risk of runs, stablecoins maintain their pegs from token holders lending out coins to leveraged traders. \cite{MaEtAl} focuses on the relationship between stablecoin runs and arbitrage centralization, finding a tradeoff between price stability and financial stability as the concentration of the arbitrageur sector varies. \cite{Bertsch} explores how the perceived quality of reserves affects the adoption of stablecoins in the presence of outside options. \cite{AhmedEtAl} study the effect that public information has on run probability, and is closely related to our work. We differ from them in a few key aspects by accounting for large sellers and analyzing uncertainty in both public and private information.

Our contribution is threefold. First, we study the effect of a large seller on selling pressure. Second, we consider the dispersion of private signals as a factor influencing stability. Third, we introduce a risk framework based on our model and use it to evaluate stablecoin fundamentals. 


On the empirical side, see \cite{LiuEtAl} which studies stablecoin flows during the Terra--Luna crash, \cite{LVN} which finds a correlation between peg efficiency and arbitrage volume, and \cite{LiaoCara} which examines the effect of stablecoin activity on the real economy. Catalini et al. offers a survey of the economics of stablecoins.

This paper is also related to the literature on global games pioneered by by \cite{CvD}, and first applied to currency attacks by \cite{MS98} and \cite{CorsettiEtAl}. Despite similarity in the reduced-form games between our model and currency attack models, the microfoundations differ. Agents in currency attack games all seek to profit from a large sale, whereas we consider the unique struture of the stablecoin market described above and assume that investors find value in holding stablecoins given issuer solvency. \cite{MS98} explore the behavior of speculators in the presence of a government that can choose to defend the peg. Similar to this paper is the work of \cite{CorsettiEtAl}, studying the behavior of a large speculator and small speculators in simultaneous- and sequential-move global games. Our model contrasts itself by relaxing the speculator's behavior to be non-strategic in the sequential-move setting. This allows us to derive comparative statics and risk decomposition results, whereas they provide numerical results and characterization of equilibria in asymptotic regimes of the information structure (for the sequential-move setting).

\section{Model}\label{sec:model}

\subsection{Complete Information Game}


\paragraph{Agents and the Market.} The stablecoin issuer's reserve asset quality is characterized by fundamentals $\theta\in\R$. There is a large seller holding a mass $\delta\in(0,1)$ of the stablecoin, and a continuum of investors with mass $1-\delta$ where each investor holds 1 stablecoin. The large seller and investors are participants in the secondary market, and sell to arbitrageurs who we assume must hold zero net inventory at the game's end. The value of the reserve assets is given by $v(\theta)=\theta$ and the stablecoin's price is initially 1.

\paragraph{Timeline.} Time is indexed by $t\in\{0,1,2,3\}$. In Period 0, Nature realizes the value of $\theta$, which is made common knowledge to all investors. In Period 1, with probability $\pi\in[0,1]$, the large seller sells his entire mass of stablecoins (for exogenous reasons). In Period 2, investors observe whether the large seller's sells or not, then each decide whether to sell their stablecoin or to continue holding on to it. Holding generates a return of $\eta>0$ conditional on the issuer's solvency.\footnote{Returns to holding could arise from investors participating in activities such as lending or staking. \cite{GortonEtAl} and \cite{MaEtAl} make similar assumptions.} In Period 3, the arbitrageur redeems all stablecoins bought from agents in Periods 1--2, possibly causing a stablecoin run. If a run occurs, then the stablecoin drops in price to 0.

\paragraph{Utilities.} Let $\lambda$ be the total mass of selling agents (including the large seller). Investors who sell always receive a payoff of 1 since the transaction occurred on the secondary market. If $\lambda<\theta$, then the issuer remains solvent, so investors who held receive a payoff of $1+\eta$ as the issuer can support par redemption at the time the game ends and holding tokens generates returns. If $\lambda\geq\theta$, then the issuer is insolvent, investors who held receive a payoff of 0 as the issuer can no longer support redemptions. 

\begin{center}
    \small
    \begin{tabular}{|c|c|c|}
        \hline
         & Issuer Solvent $(\lambda<\theta)$ & Issuer Insolvent $(\lambda\geq\theta)$ \\
        \hline
        Action $=$ Sell & $1$ & $1$ \\
        \hline
        Action $=$ Hold & $1+\eta$ & $0$ \\
        \hline
    \end{tabular}
    \captionof{table}{\small An investor's utility in the game of complete information.}
    \label{fig:timeline}
\end{center}

\paragraph{Equilibrium.} Observe that holding strictly dominates selling if the issuer is solvent and selling strictly dominates holding if the issuer is insolvent. Then:

\begin{itemize}
    \item Suppose that there is no large sale. If $\theta>1-\delta$, then the issuer is solvent even if all investors sell, so the only equilibrium strategy is for all investors to hold. If $\theta\leq1-\delta$, then there exists a critical proportion $\lambda^\star_0=\theta/(1-\delta)$. For $\lambda<\lambda^\star_0$, selling investors finds it more profitable to hold; for $\lambda>\lambda^\star_0$, holding investors finds it more profitable to sell. It follows that the only equilibrium strategies are (i) for all investors to hold or (ii) for all investors to sell.
    \item Suppose that a large sale occurs. If $\theta\leq\delta$, then the issuer is insolvent even if no investors sell, so the equilibrium strategy is where all investors will sell. If $\theta\in(\delta,1]$, then by similar reasoning, the equilibrium strategies are (i) for all investors to hold or (ii) for all investors to sell. If $\theta>1$, then then the issuer is solvent even if all investors sell, so the only equilibrium strategy in this case is for all investors to hold.
\end{itemize}

Since $v$ is increasing, there exist thresholds for $\theta$ such that the ``all-sell'' and ``all-hold'' equilibria are unique given a large sale and the ``all-hold'' equilibrium given no large sale. For the other cases, the existence of multiple equilibria reflects the classic self-fulfilling property of bank runs and currency attacks where investors hold only if they believe that other investors will hold, and vice versa. In the common knowledge game, the large seller exerts selling pressure on the investors: the region for $\theta$ where the all-hold equilibrium is unique shrinks, and a region for $\theta$ where the all-sell equilibrium is unique is present (whereas such as region is not present in the absence of a large sale).

\begin{center}
    \small
    \begin{tabular}{|c|c|c|c|c|}
        \hline
         & $\theta\in(-\infty,\delta]$ & $\theta\in(\delta,1-\delta]$ & $\theta\in(1-\delta,1]$ & $\theta>1$ \\
        \hline
        Large Sale Occurs & $\lambda^\star_I=1$ & $\lambda^\star_I\in\{0,1\}$ & $\lambda^\star_I\in\{0,1\}$ & $\lambda^\star_I=0$ \\
        \hline
        No Large Sale Occurs & $\lambda^\star_I\in\{0,1\}$ & $\lambda^\star_I\in\{0,1\}$ & $\lambda^\star_I=0$ & $\lambda^\star_I=0$ \\
        \hline
    \end{tabular}
    \captionof{table}{\centering \small Possible equilibrium proportions $\lambda^\star_I$ of investors (not all agents) who sell in the game of complete information, depending on whether a large sale occurs or not.}
    \label{fig:timeline}
\end{center}

\subsection{Global Game}

To resolve the issue of multiple equilibria, we modify the game into a \textit{global game} where each agent in the continuum receives a private signal about $\theta$. The agents and market remain the same. The timeline and utilities are modified as follows.

\paragraph{Timeline (Global Game).} Time is indexed by $t\in\{0,1,2,3\}$. In Period 0, Nature draws a fundamental value $\theta\sim N(\mu,\sigma^2)$. Given the realized value of $\theta$, investor $i$ receives a signal $x_i=\theta+\sigma\varepsilon_i$ where $\varepsilon_i\sim N(0,\sigma_x^2)$. From this perspective, the posterior distribution of investor $i$ is $\theta|x_i\sim N(\mu_p(x_i),\sigma_p)$ where
\begin{gather*}
    \mu_p(x_i) = \frac{\sigma^2x_i+\sigma_x^2\mu}{\sigma^2+\sigma_x^2}, \quad \sigma_p = \frac{\sigma^2\sigma_x^2}{\sigma^2+\sigma_x^2}.
\end{gather*}

In Period 1, with probability $p\in[0,1]$, the large seller sells his entire mass of stablecoins. In Period 2, investors observe whether the large sale occurs or not, then each decide whether to sell their stablecoin or to continue holding on to it based on their signal. Conditional on the issuer's solvency, holding tokens generates a return of $\eta>0$. In Period 3, the arbitrageur redeems all stablecoins bought from agents in Periods 1 and 2, possibly causing a stablecoin run. If a run occurs, then the stablecoin drops in price to 0.

\begin{remark}
A natural modeling question is whether the large seller’s decision should be treated as endogenous rather than exogenously specified as it is in our case. Compared to \cite{CorsettiEtAl}, which considers an endogenous large seller, this model produces the same key insight (of increased selling pressure) without introducing multiplicity of equilibria in the global game. Additionally, it focuses the analysis on the \textit{event} of a large redemption occurring without committing to a specific motive, which may vary (such as speculation, liquidity needs, risk management). The resulting framework allows for uniqueness, tractability, and interpretable comparative statics.
\end{remark}

\paragraph{Utilities (Global Game).} For fundamentals $\theta$ and the total mass $\lambda$ of redeeming agents, an investor's utility remains the same. However, they now evaluate the \textit{expected} utility with respect to their posterior over $\theta$ given $\bar x_i$. Moreover, $\lambda$ may also depend on $\theta$. We work with an investor's expected \textit{payoff differential} between selling and holding. For an investor receiving a signal of $x_i$, this is
\begin{align*}
    \Delta(x_i)=\E_{\theta\sim F_{\mu_p(x_i),\sigma_p}}[\1_{\{\lambda(\theta)\geq \theta\}}-\eta\cdot\1_{\{\lambda(\theta)<\theta\}}]
\end{align*}
where $F_{\mu,\sigma}$ is the distribution of a $N(0,\sigma^2)$ random variable.

\paragraph{Equilibrium (Global Game).} We search for equilibria in \textit{switching strategies} where investors redeem if and only if their signal is below a threshold value. For a switching strategy specified by $\bar x$, the proportion of investors who redeem when Nature's realization of fundamentals is $\theta$, denoted by $\lambda(\theta;\bar x)$, is
\begin{align*}
    \lambda_I(\theta;\bar x) = F_{\theta,\sigma_x}(\bar x)=1-F_{\bar x,\sigma_x}(\theta)
\end{align*}
due to the symmetry the normal distribution. For fixed $\bar x$, as $\lambda_I$ is decreasing in $\theta$ and $v$ is increasing in $\theta$, the not-too-large assumption implies that there exists unique values of $\bar\theta_0(\bar x)$ and $\bar\theta_1(\bar x)$ where
\begin{gather*}
    (1-\delta)\cdot\lambda_I(\bar\theta_0(\bar x);\bar x)=\bar\theta_0(\bar x) \\
    \delta+(1-\delta)\cdot\lambda_I(\bar\theta_1(\bar x);\bar x)=\bar\theta_1(\bar x).
\end{gather*}
These are the thresholds where the issuer is solvent when the large sale occurs and does not occur, respectively. For $\theta$ values below $\bar\theta_0(\bar x)$ or $\bar\theta_1(\bar x)$, the issuer is insolvent, and for values above $\bar\theta_0(\bar x)$ or $\bar\theta_1(\bar x)$, the issuer is solvent. An investor's expected payoff differential who receives signal $\theta_i$ and follows switching strategy $\bar x$ can then be written as
\begin{gather*}
    \Delta_0(x_i;\bar x) = \int_{-\infty}^{\bar\theta_0(\bar x)}dF_{\mu_p(x_i),\sigma_p}(\theta)-\int_{\bar\theta_0(\bar x)}^{\infty}\eta\,dF_{\mu_p(x_i),\sigma_p}(\theta) \\[0.25\baselineskip] 
    \Delta_1(x_i;\bar x) = \int_{-\infty}^{\bar\theta_1(\bar x)}dF_{\mu_p(x_i),\sigma_p}(\theta)-\int_{\bar\theta_1(\bar x)}^{\infty}\eta\,dF_{\mu_p(x_i),\sigma_p}(\theta)
\end{gather*}
where the 0 and 1 signify the cases where a large sale occurs and does not occur, respectively.

For a switching strategy $\bar x$ to be an equilibrium, the investor who receives a signal of exactly $\bar x$ must be indifferent between selling and holding; thus their payoff differential
\begin{align*}
    \bar{\Delta}_a(\bar x)\equiv\Delta_a(\bar x;\bar x)=F_{\mu_p(\bar x),\sigma_p}(\bar\theta_a(\bar x))-\eta(1-F_{\mu_p(\bar x),\sigma_p}(\bar\theta_a(\bar x)))
\end{align*}
must be zero for $\alpha\in\{0,1\}$. Moreover, investors receiving signals below $\bar x$ must have a positive payoff differential (prefer to sell) and investors receiving signals above $\bar x$ must have a negative payoff differential (prefer to hold). As $F_{\mu_p(x_i),\sigma_p}(\bar x)$ is decreasing in $x_i$, this condition is satisfied. To establish the uniqueness of an equilibrium switching strategy, it suffices to show that $\bar{\Delta}_a(\bar x)$ crosses zero exactly once over $\bar x\in\R$. An equilibrium in the global game is then specified by $(x^\star_0,x^\star_1)$ where $x^\star_0$ and $x^\star_1$ are equilibrium switching strategies for investors conditional on a large sale occurring and not occurring, respectively. For the remainder of this paper, we also assume that $\eta$, the returns from holding stablecoins, and $\sigma_x$, the dispersion of investors' private signals, are not too large.

\begin{assumption}
We assume that $\eta<1$ and $(1-\delta)\sigma_x<\sqrt{2\pi}\sigma^3$.
\end{assumption}

\begin{proposition}
Under Assumption 1, there exists a unique equilibrium $(x^\star_0,x^\star_1)$ in switching strategies of the global game in which $x^\star_0<x^\star_1$.
\end{proposition}

The above proposition establishes the existence of a unique equilibrium in the global game and asserts that the switching threshold when the large sale is greater than that when the large sale does not sell. This difference can be interpreted as an increase in selling pressure in the presence of a large sale as investors who receive signals between $x^\star_0$ and $x^\star_1$, who would not sell in the absence of a large sale, now find it optimal to sell. 

Given an equilibrium switching strategy $(x^\star_0,x^\star_1)$, it follows that when is there is no large sale and the realized value of $\theta$ falls below $\bar\theta_0(x^\star_0)$, the threshold at which the proportion of sellers is equal to the normalized value of reserves in the absence of a large sale, the stablecoin will run . Similarly, when there is a large sale and the realized value of $\theta$ falls below $\bar\theta_1(x^\star_1)$, the threshold at which the proportion of sellers is equal to the normalized value of reserves in the presence of a large sale, the stablecoin will run as well. We show in the appendix that $\bar\theta_0(x^\star_0)<\bar\theta_0(x^\star_1)<\bar\theta_1(\theta^\star_1)$, meaning that the threshold of $\theta$ at which a run occurs is larger given a large sale than given no large sale.


\section{Results}

In the previous section, we determined basic results about equilibrium in the global game and the effect of the large seller on the selling behavior of the investors. We now focus our attention to the probability of a stablecoin run implied by an equilibrium switching strategy. The ex-ante run probability $R$ evaluated \textit{before} Nature draws a value of $\theta$ from the $ N(\mu,\sigma^2)$ distribution is
\begin{align*}
    R=(1-p)\cdot F_{\mu,\sigma}(\bar\theta_0(x_0^\star))+p\cdot F_{\mu,\sigma}(\bar\theta_1(x_1^\star)).
\end{align*}
The first term captures the probability of run based on fundamentals alone. The second term captures the run probability induced by a large sale. We refer to these terms as the \textit{collateral risk} (denoted $R_0$) and the \textit{large sale risk} (denoted $R_1$), respectively, and given by the following:
\vspace{-0.1em}
\begin{gather*}
    R_0 = (1-p)\cdot F_{\mu,\sigma}(\bar\theta_0(x_0^\star)), \\
    R_1 = p\cdot F_{\mu,\sigma}(\bar\theta_1(x_1^\star)).
\end{gather*}
This decomposition allows us to conceptualize how much two different types of risk present in the stablecoin ecosystem contribute to the overall run probability of an issuer. Although we will have always have $F_{\mu,\sigma}(\bar\theta_0(x^\star_0))<F_{\mu,\sigma}(\bar\theta_1(x^\star_1))$, the value of $p$ affects the magnitudes of $R_0$ and $R_1$, as seen in Figure 1. 


\vspace{0.1em}
\begin{center}
    \tikzset{every picture/.style={line width=0.75pt}} 
    
    \begin{tikzpicture}[x=0.75pt,y=0.75pt,yscale=-1,xscale=1]
    
    \draw [color={rgb, 255:red, 208; green, 2; blue, 27 }  ,draw opacity=1 ]   (100.33,120.67) .. controls (170.33,120.67) and (149.33,229) .. (220.33,229) ;
    \draw [color={rgb, 255:red, 208; green, 2; blue, 27 }  ,draw opacity=1 ]   (220.33,229) -- (307.33,229.33) ;
    \draw [color={rgb, 255:red, 208; green, 2; blue, 27 }  ,draw opacity=1 ]   (80.33,120.33) -- (100.33,120.67) ;
    \draw [color={rgb, 255:red, 74; green, 144; blue, 226 }  ,draw opacity=1 ]   (141.33,190.33) .. controls (211.33,190.33) and (188.33,228) .. (259.33,228) ;
    \draw [color={rgb, 255:red, 74; green, 144; blue, 226 }  ,draw opacity=1 ]   (256.33,228) -- (307.33,228) ;
    \draw [color={rgb, 255:red, 74; green, 144; blue, 226 }  ,draw opacity=1 ]   (80.33,190.33) -- (141.33,190.33) ;
    \draw    (77,238) -- (334.33,237.67) ;
    \draw [shift={(336.33,237.67)}, rotate = 179.93] [color={rgb, 255:red, 0; green, 0; blue, 0 }  ][line width=0.75]    (10.93,-3.29) .. controls (6.95,-1.4) and (3.31,-0.3) .. (0,0) .. controls (3.31,0.3) and (6.95,1.4) .. (10.93,3.29)   ;
    \draw    (77,238) -- (77.33,79.67) ;
    \draw [shift={(77.33,77.67)}, rotate = 90.12] [color={rgb, 255:red, 0; green, 0; blue, 0 }  ][line width=0.75]    (10.93,-3.29) .. controls (6.95,-1.4) and (3.31,-0.3) .. (0,0) .. controls (3.31,0.3) and (6.95,1.4) .. (10.93,3.29)   ;
    \draw  [dash pattern={on 0.84pt off 2.51pt}]  (157,171) -- (157,236) ;
    \draw  [dash pattern={on 0.84pt off 2.51pt}]  (199.33,212) -- (199.33,236) ;
    
    \draw (58,185.4) node [anchor=north west][inner sep=0.75pt]  [font=\small]  {$p$};
    \draw (37,112.4) node [anchor=north west][inner sep=0.75pt]  [font=\small]  {$1-p$};
    \draw (88,172.4) node [anchor=north west][inner sep=0.75pt]  [font=\small]  {$R_1$};
    \draw (88,102.4) node [anchor=north west][inner sep=0.75pt]  [font=\small]  {$R_0$};
    \draw (140,242.4) node [anchor=north west][inner sep=0.75pt]  [font=\small]  {$\bar\theta_0(x^\star_0)$};
    \draw (183,242.4) node [anchor=north west][inner sep=0.75pt]  [font=\small]  {$\bar\theta_1(x^\star_1)$};
    \draw (62,220.4) node [anchor=north west][inner sep=0.75pt]  [font=\small]  {$0$};
    \end{tikzpicture}
    \captionof{figure}{\centering \small A diagram 
    of collateral risk (red) and large sale risk (blue). The horizontal axis represents the value of fundamentals and the vertical axis represents probabilities.}
\end{center}

\begin{proposition}
Collateral risk decreases and large sale risk increases as the large seller's size grows. Specifically, $dR_0/d\delta<0$ and $dR_1/d\delta>0$.
\end{proposition}

The above proposition states that collateral risk is decreasing and large sale risk is increasing in the size of the large seller. As the large seller becomes more influential relative to the other investors, when no large sale occurs, a smaller mass of investors make runs less likely. Conversely, when there is a large sale, the additional influence of the large seller drives up the run probability. How this affects the total ex-ante run probability depends on the value of $p$ and the relative magnitude of the rates of change.

We now consider the effect of information uncertainty on the risk components and overall run probability. Two parameters control the uncertainty of information in our model: the scale of the common prior $\sigma$ and the scale of dispersion in private signals $\sigma_x$. Shocks that are known to all investors affect $\sigma$, while factors related to the heterogeneity of investors' beliefs affect $\sigma_x$. 

An example of a shock to the common prior's variance could be an issuer's disclosure of reserve assets. Aspects influencing the magnitude of the shock include frequency, verifiability, and content of the disclosure. Elements affecting the dispersion of investors' beliefs include information asymmetry and the nature of collateral reserves. For example, as an opaque stablecoin issuer covertly recollateralizes, information about the transfers could be leaked to select market participants, widening the information gap between more and less informed investors. Having stablecoins collateralized across different assets or asset classes can also drive up the scale of signal dispersion, as variation in investors' beliefs is amplified by the variety of collateral reserves. 

To facilitate analysis, we use precision parameters $\tau=\sigma^{-2}$ and $\tau_x=\sigma_x^{-2}$. Proposition 3.2 characterizes the comparative statics of risk components with respect to the precision of commonly known information, and Proposition 3.3 does the same but with respect to the precision of private signals.

\begin{proposition}
If the quality of fundamentals is sufficiently high (low), then collateral risk and large sale risk decrease (increase) as public knowledge is more precise. Specifically, for $a\in\{0,1\}$:
\begin{gather*}
    \mu>\bar\theta_a(x^\star_a)+\frac{2\tau+\tau_x}{3\tau_x\sqrt{\tau+\tau_x}}\cdot F^{-1}(1/(1+\eta)) \implies \frac{dR_a}{d\tau}<0, \\[0.25\baselineskip]
    \mu<\bar\theta_a(x^\star_a) \implies \frac{dR_a}{d\tau}>0.
\end{gather*}
\end{proposition}

When varying the dispersion of the common prior, Proposition 3.2 reveals that for each risk component, there exists upper and lower thresholds at which the risk component increases in $\tau$ for values of $\theta$ below the lower threshold and decreases in $\tau$ for values of $\theta$ above the upper threshold. In other words, when quality of fundamentals is poor, more perceived uncertainty about the value of fundamentals from the investor's point of view helps reduce the risk of a run. This is because low uncertainty concentrates investors' signals around a poor fundamental quality, meaning that an individual investor assesses that other investors have similar pessimistic beliefs about the state of reserves as themselves. Conversely, when the quality of fundamentals is good, more uncertainty increases the risk of a run. The economic reasoning for this is similar: low uncertainty concentrates the signals of investors around a good fundamental quality, so an investor gauges that other investors have similar optimistic beliefs.


\begin{center}
    
\tikzset{every picture/.style={line width=0.75pt}} 

\begin{tikzpicture}[x=0.75pt,y=0.75pt,yscale=-1,xscale=1]

\draw    (180.33,240) -- (488.33,240) ;
\draw [shift={(490.33,240)}, rotate = 180] [color={rgb, 255:red, 0; green, 0; blue, 0 }  ][line width=0.75]    (10.93,-3.29) .. controls (6.95,-1.4) and (3.31,-0.3) .. (0,0) .. controls (3.31,0.3) and (6.95,1.4) .. (10.93,3.29)   ;
\draw  [dash pattern={on 4.5pt off 4.5pt}]  (180.33,195) -- (488,195) ;
\draw    (270,230) -- (270,250) ;
\draw    (400,230) -- (400,250) ;
\draw    (290,230) -- (290,250) ;
\draw    (380,230) -- (380,250) ;
\draw  [dash pattern={on 4.5pt off 4.5pt}]  (270,198) -- (270,229) ;
\draw  [dash pattern={on 4.5pt off 4.5pt}]  (290,198) -- (290,228) ;
\draw  [dash pattern={on 4.5pt off 4.5pt}]  (380,152) -- (380,193) ;
\draw  [dash pattern={on 4.5pt off 4.5pt}]  (400,152) -- (400,192) ;

\draw (138,157.4) node [anchor=north west][inner sep=0.75pt]  [font=\small]  {$\frac{\partial R_1}{\partial \tau }$};
\draw (138,204.4) node [anchor=north west][inner sep=0.75pt]  [font=\small]  {$\frac{\partial R_0}{\partial \tau }$};
\draw (268.31,253.6) node [anchor=north west][inner sep=0.75pt]  [font=\small,rotate=-30]  {$\overline{\theta }_{0}\!\left( x_{0}^{\star }\right)$};
\draw (383,214.4) node [anchor=north west][inner sep=0.75pt]    {$-$};
\draw (219,214.4) node [anchor=north west][inner sep=0.75pt]    {$+$};
\draw (438,166.4) node [anchor=north west][inner sep=0.75pt]    {$-$};
\draw (274,166.4) node [anchor=north west][inner sep=0.75pt]    {$+$};
\draw (378.39,253.6) node [anchor=north west][inner sep=0.75pt]  [font=\small,rotate=-30]  {$\overline{\theta }_{1}\!\left( x_{1}^{\star }\right)$};
\draw (296.68,250.81) node [anchor=north west][inner sep=0.75pt]  [font=\small,rotate=-30]  {$\overline{\theta }_{0}\!\left( x_{0}^{\star }\right) +c$};
\draw (408.53,250.81) node [anchor=north west][inner sep=0.75pt]  [font=\small,rotate=-30]  {$\overline{\theta }_{1}\!\left( x_{1}^{\star }\right) +c$};

\end{tikzpicture}
\captionof{figure}{\centering \small Diagram of how collateral risk and large sale risk change with respect to the precision of the common prior. The horizontal axis represents the prior mean $\mu$, and $c=\frac{2\tau+\tau_x}{3\tau_x\sqrt{\tau+\tau_x}}\cdot F^{-1}(1/(1+\eta)) \implies \frac{dR_a}{d\tau}<0$.}
\end{center}

Figure 2 illustrates how each risk component varies with respect to changes in the common prior's uncertainty, given different values of $\theta$. By having a large seller in our model, we can identify three regions of interest for the fundamental space: (i) where more precision increases both types of risk, (ii) where more precision reduces both types of risk, and (iii) where more precision increases large sale risk but decreases collateral risk. Such a region is present when high precision concentrates the signals of investors around values that are safe from runs when no large sale occurs, but are vulnerable to runs when a large sale occurs. Depending on the value of $p$, opaqueness may reduce the overall risk for an issuer when the prior mean falls within this range.

Figure 3 provides a visual explanation of why this occurs. While both prior distributions in the figure has means between $\bar\theta_0(x^\star_0)$ and $\bar\theta_1(x^\star_1)$, the brown line, representing a prior of higher uncertainty, has more density below $\bar\theta_0(x^\star_0)$ but less density below $\bar\theta_1(x^\star_1)$ than the orange line, which is a prior of lower uncertainty. Since the realization of $\theta$ from the prior and whether there is a large sale determines if a run occurs, this difference translates to the relative difference in risk component values as well.

\vspace{0.1em}
\begin{center}

\tikzset{every picture/.style={line width=0.75pt}} 

\begin{tikzpicture}[x=0.75pt,y=0.75pt,yscale=-1,xscale=1]

\draw    (180.33,240) -- (458.33,240) ;
\draw [shift={(460.33,240)}, rotate = 180] [color={rgb, 255:red, 0; green, 0; blue, 0 }  ][line width=0.75]    (10.93,-3.29) .. controls (6.95,-1.4) and (3.31,-0.3) .. (0,0) .. controls (3.31,0.3) and (6.95,1.4) .. (10.93,3.29)   ;
\draw    (270,230) -- (270,250) ;
\draw    (361,230) -- (361,250) ;
\draw  [dash pattern={on 4.5pt off 4.5pt}]  (270,149) -- (270,226) ;
\draw  [dash pattern={on 4.5pt off 4.5pt}]  (361,149) -- (361,226) ;
\draw [color={rgb, 255:red, 139; green, 87; blue, 42 }  ,draw opacity=1 ]   (225,235) .. controls (271,235) and (269,179) .. (314,179) ;
\draw [color={rgb, 255:red, 139; green, 87; blue, 42 }  ,draw opacity=1 ]   (405,235) .. controls (359,234) and (362,179) .. (314,179) ;
\draw [color={rgb, 255:red, 139; green, 87; blue, 42 }  ,draw opacity=1 ]   (190,235) -- (225,235) ;
\draw [color={rgb, 255:red, 139; green, 87; blue, 42 }  ,draw opacity=1 ]   (405,235) -- (440,235) ;
\draw [color={rgb, 255:red, 245; green, 166; blue, 35 }  ,draw opacity=1 ]   (250,237) .. controls (291,237) and (289,157) .. (314,157) ;
\draw [color={rgb, 255:red, 245; green, 166; blue, 35 }  ,draw opacity=1 ]   (380,237) .. controls (340,237) and (342,157) .. (314,157) ;
\draw [color={rgb, 255:red, 245; green, 166; blue, 35 }  ,draw opacity=1 ]   (190,237) -- (250,237) ;
\draw [color={rgb, 255:red, 245; green, 166; blue, 35 }  ,draw opacity=1 ]   (380,237) -- (440,237) ;

\draw (268.31,253.56) node [anchor=north west][inner sep=0.75pt]  [font=\small,rotate=-30]  {$\overline{\theta }_{0}\!\left( x_{0}^{\star }\right)$};
\draw (359.39,253.6) node [anchor=north west][inner sep=0.75pt]  [font=\small,rotate=-30]  {$\overline{\theta }_{1}\!\left( x_{1}^{\star }\right)$};
\end{tikzpicture}
\captionof{figure}{\centering \small Comparison of priors with different precisions with critical run thresholds marked. The horizontal axis represents fundamental values $\theta$.}
\end{center}

\begin{proposition}
If the quality of fundamentals is sufficiently high (low), then collateral risk and large sale risk increase (decrease) as private signals are more precise. Specifically, for $a\in\{0,1\}$:
\begin{gather*}
    \mu>\bar\theta_a(x^\star_a)+\frac{F^{-1}(1/(1+\eta))}{\sqrt{\tau+\tau_x}} \implies \frac{dR_a}{d\tau_x}>0, \\[0.25\baselineskip]
    \mu<\bar\theta_a(x^\star_a) \implies \frac{dR_a}{d\tau_x}<0.
\end{gather*}
\end{proposition}

We now look at the risk components as private signal dispersion varies. Contrary to Proposition 3.2, Proposition 3.3 states that for each component, there exists a single threshold at which the risk component \textit{decreases} in $\tau_x$ for values of $\theta$ below the threshold and \textit{increases} in $\tau_x$ for values of $\theta$ above the threshold. To elaborate, when the quality of fundamentals is good, more dispersion in private signals reduces the run risk, whereas when fundamentals are poor, more varied private signals increase the risk of run. This seemingly paradoxical implication that more uncertainty improves financial stability in good states can be explained by the connection between global games and risk dominance.

\cite{CvD} show that a risk dominant equilibrium is selected in global games. When the variance of private signals is low, the global game approaches the common knowledge game where there are multiple equilibria, one payoff dominant and one risk dominant. As the variance of private signals increases, the investors become more uncertain about each other's signals and thus actions, heightening the degree of risk dominance. Intuitively,  the ``safe'' option is to hold when fundamentals are good, and sell when fundamentals are poor. Thus, Propositions 3.2 and 3.3 contrast as the common prior's variance drives investors' uncertainty about the stablecoin reserves whereas the private signal's variance drives investors' uncertainty about each other's actions.

\vspace{0.1em}

\begin{center}
    
\tikzset{every picture/.style={line width=0.75pt}} 

\begin{tikzpicture}[x=0.75pt,y=0.75pt,yscale=-1,xscale=1]

\draw    (180.33,240) -- (488.33,240) ;
\draw [shift={(490.33,240)}, rotate = 180] [color={rgb, 255:red, 0; green, 0; blue, 0 }  ][line width=0.75]    (10.93,-3.29) .. controls (6.95,-1.4) and (3.31,-0.3) .. (0,0) .. controls (3.31,0.3) and (6.95,1.4) .. (10.93,3.29)   ;
\draw  [dash pattern={on 4.5pt off 4.5pt}]  (180.33,195) -- (488,195) ;
\draw    (270,230) -- (270,250) ;
\draw    (400,230) -- (400,250) ;
\draw    (290,230) -- (290,250) ;
\draw    (380,230) -- (380,250) ;
\draw  [dash pattern={on 4.5pt off 4.5pt}]  (270,198) -- (270,229) ;
\draw  [dash pattern={on 4.5pt off 4.5pt}]  (290,198) -- (290,228) ;
\draw  [dash pattern={on 4.5pt off 4.5pt}]  (380,152) -- (380,193) ;
\draw  [dash pattern={on 4.5pt off 4.5pt}]  (400,152) -- (400,192) ;

\draw (138,157.4) node [anchor=north west][inner sep=0.75pt]  [font=\small]  {$\frac{\partial R_1}{\partial \tau }$};
\draw (138,204.4) node [anchor=north west][inner sep=0.75pt]  [font=\small]  {$\frac{\partial R_0}{\partial \tau }$};
\draw (268.31,253.6) node [anchor=north west][inner sep=0.75pt]  [font=\small,rotate=-30]  {$\overline{\theta }_{0}\!\left( x_{0}^{\star }\right)$};
\draw (383,214.4) node [anchor=north west][inner sep=0.75pt]    {$+$};
\draw (219,214.4) node [anchor=north west][inner sep=0.75pt]    {$-$};
\draw (438,166.4) node [anchor=north west][inner sep=0.75pt]    {$+$};
\draw (274,166.4) node [anchor=north west][inner sep=0.75pt]    {$-$};
\draw (378.39,253.6) node [anchor=north west][inner sep=0.75pt]  [font=\small,rotate=-30]  {$\overline{\theta }_{1}\!\left( x_{1}^{\star }\right)$};
\draw (296.68,250.81) node [anchor=north west][inner sep=0.75pt]  [font=\small,rotate=-30]  {$\overline{\theta }_{0}\!\left( x_{0}^{\star }\right) +c$};
\draw (408.53,250.81) node [anchor=north west][inner sep=0.75pt]  [font=\small,rotate=-30]  {$\overline{\theta }_{1}\!\left( x_{1}^{\star }\right) +c$};

\end{tikzpicture}
\captionof{figure}{\centering \small Diagram of how collateral risk and large sale risk change with respect to the precision of the common prior. The horizontal axis represents the prior mean $\mu$, and $c=\frac{F^{-1}(1/(1+\eta))}{\sqrt{\tau+\tau_x}}$.}
\end{center}

Similarly to Figure 2, Figure 4 shows how the risk components vary with respect to changes in the dispersion of private signals. There exist regions where (i) both risk components increase as signals are more precise, (ii) both risk components decrease as signals are more precise, and (iii) where collateral risk decreases but large sale risk increases as signals are more precise. Regions (i) and (iii) suggest that depending on the values of $\mu$ and $p$, more heterogeneity in investors' beliefs may be beneficial for the stablecoin issuer's financial stability. This heterogeneity can arise from a naturally diverse investor base, or by the stablecoin's nature (e.g. information leaks or multi-collateralization).

We can also examine the share of the total run probability that each risk component contributes towards. We denote the \textit{share of collateral risk} and the \textit{share of large sale risk} as $S_0$ and $S_1$, respectively, which are given by
\begin{gather*}
    S_0 = \frac{R_0}{R_0+R_1}, \quad S_1 = \frac{R_1}{R_0+R_1}.
\end{gather*}
As an immediate corollary of Proposition 3.1, we have that the share of large sale risk increases in the size of the large seller. For the regions implied by Propositions 3.2 and 3.3 where one type of risk increases and the other decreases as information is more precise, we can deduce that the share of large sale risk decreases in the variance of the common prior and increases in the variance of private signals for an intermediate range of $\mu$. As the average quality of fundamentals itself, captured by $\mu$, changes, we have the following result.

\begin{proposition}
The share of large sale risk increases in the average quality of fundamentals. Specifically, $dS_1/d\mu>0$.
\end{proposition} 

When varying $\mu$ and fixing all other parameters, the share of large sale risk increases in the average quality of fundamentals. This makes intuitive sense as collateral risk is a very present issue when fundamentals are poor, while when fundamentals are good, the risk of running based on fundamentals alone decreases. Thus, in good states, much of the run risk comes from a large seller exerting selling pressure on the other investors.

\section{Discussion}

\subsection{Case Studies}

We begin the discussion by relating our model's results and implications with the case studies described in the introduction.
\begin{itemize}
    \item \textbf{Circle and the Silicon Valley Bank Collapse.} Since USDC is a fiat-backed stablecoin whose issuer, Circle, makes frequent public disclosures about reserves which are generally trusted, USDC would correspond to a stablecoin with a low prior variance. Our model predicts good financial stability when fundamentals are good and poor financial stability when fundamentals are poor, relative to coins of a similar reserve structure but different levels of transparency. During SVB's collapse, when Circle disclosed that SVB held a portion of their reserves, USDC depegged shortly after. Upon the announcement by the Federal Deposit Insurance Corporation (FDIC) that SVB would be bailed out, USDC repegged not long after. In this case, as a consequency of transparency, collateral risk is a major determinant of financial stability, as opposed to large sale risk.
    \item \textbf{Tether's Disclosures and USDT Depegging Events.} For Tether, disclosures of reserves are less frequent and less verifiable due to operating outside of a regulatory framework, resulting in a higher prior variance. Compared to more transparent stablecoins, there is more exposure to run risk when fundamentals are good, but less exposure to large sale risk when fundamentals are lower but not too low (see the intermediate interval in Figure 2). Additionally, the asymmetric information generated by Tether's transfers and recollateralizations corresponds to a large dispersion in private signals, which may additionally reduce run risk when fundamentals are good. These reasons could explain why despite the opaqueness and occasional depegging driven by large sales, Tether repegs in a short amount of time, avoiding more substantial depegging incidents or a full collapse. 
    \item \textbf{Terra--Luna Collapse.} Endogenously-backed stablecoins like UST tend to have high prior variances due to their self-referential nature. For example, \cite{AhmedEtAl} find empirical evidence that Luna, the cryptoasset backing TerraUSD, had a non-constant conditional volatility, leading to unconditional fat tails. Additionally, our model suggests that single-collateralized stablecoins, including UST, should generate less heterogeneous beliefs and thus more precise private signals, heightening large sale risk when fundamentals are good. As the collapse of UST and Luna started with a large sale, possibly of speculative nature, it could be that a combination of relatively large prior variance and precise private signals contributed to the full collapse.
    \item \textbf{GENIUS Act.} The GENIUS Act's requirements of full backing and transparent reporting significantly changes the environment in which US-based stablecoin issuers and investors interact. In our model's terms, the act requires (i) stablecoin fundamentals to be consistently good and (ii) public information to be very precise. This act resolves the issues with Circle, which held risky cash deposits, by requiring high quality reserves, as well as the issues with Tether, which raised doubts with its dubious disclosure policies. 
\end{itemize}

\subsection{Information Design and Regulation}

In our framework, greater dispersion in private signals can be a stabilizing force when reserve fundamentals are sufficiently strong, i.e.\ stablecoin issuers can benefit from a certain degree of heterogeneity in how investors interpret information. In practice, this does not mean obfuscation or misinformation, but rather design choices that naturally generate diverse views: cultivating a broad and heterogeneous user base (retail, DeFi protocols, market-makers, institutions), holding a diversified but fundamentally safe reserve portfolio, and communicating rich but non‐coarse disclosures that different investors process with different models.

The GENIUS Act, however, pushes in the opposite direction along several margins: by mandating conservative, simple reserve structures and frequent disclosures of reserve composition and redemption policies, it sharply increases the precision of public information, thus reducing room for informational heterogeneity that comes purely from opacity or selective disclosure. In a post-GENIUS environment, an issuer cannot “engineer” private signal dispersion by withholding or shading information; any attempt to do so would run into compliance and enforcement risk. Thus, US-based stablecoin issuers could be less capital efficient than offshore issuers not subject to the same regulations, who might hold riskier reserves or have opaque disclosures, but can effectively create stability via information design.

Our model implies that GENIUS tilts the information structure toward a regime where public information dominate and runs, when they occur, are more tightly tied to genuine deterioration in fundamentals. Residual dispersion in private signals coming from heterogeneous agents can still provide a buffer against self-fulfilling runs in good states, but is now a second-order issue. This highlights an important tension in the GENIUS framework: while it improves auditability and aligns runs more closely with fundamentals, it may do so at the cost of capital efficiency by mandating exclusively dollar-equivalent assets.

A more careful regulatory approach could consider alternative frameworks that preserve ex post auditability without making full public disclosure the default, such as continuous supervisory audit rights or cryptographic proof-of-reserves mechanisms that verify solvency without revealing custodial counterparties. While such approaches are likely second-order relative to the binding constraint on reserve composition, they illustrate how information policy, not just asset restrictions, shapes the competitive landscape between regulated and offshore stablecoin issuers.




\section{Conclusion}

Reserve assets backing a stablecoin and the risks facing a stablecoin issuer can vary. With a model capturing two common types of risk, one from large sales and the other from collateral shocks, we find that large sales exert selling pressure on investors and show how uncertainty in information about stablecoin reserves influences run risk. Our model has implications consistent with real-world depegging events, and suggests that more uncertainty may improve an issuer's financial stability in certain contexts. A careful understanding of these risks and the role that policies have on them is crucial when addressing the future trajectory and regulatory challenges surrounding stablecoins.

\bibliographystyle{plainurl}
\bibliography{stablecoin}

@article{AhmedEtAl,
Author = {Ahmed, Rashad and Aldasoro, Iñaki and Duley, Chanelle},
Title = {{Public information and stablecoin runs}},
Journal = {BIS Working Paper},
Volume = {1164},
Year = {2024},
Month = {},
Pages = {},
DOI = {},
URL = {}
}

@article{CorsettiEtAl,
Author = {Corsetti, Giancarlo and Dasgupta, Amil and Morris, Stephen and Shin, Hyun Song},
Title = {{Does One Soros Make a Difference? A Theory of Currency Crises with Large and Small Traders}},
Journal = {Review of Economic Studies},
Volume = {71},
Year = {2002},
Month = {},
Pages = {87--113},
DOI = {},
URL = {}
}

@article{GortonEtAl,
Author = {Gorton, Gary B. and Klee, Elizabeth C. and Ross, Chase P. and Ross, Sharon Y. and Vardoulakis, Alexandros P.},
Title = {{Leverage and Stablecoin Pegs}},
Journal = {Journal of Financial and Quantitative Analysis},
Volume = {Forthcoming},
Year = {2022},
Month = {},
Pages = {},
DOI = {},
URL = {}
}

@article{MaEtAl,
Author = {Ma, Yiming and Zeng, Yao and Zhang, Anthony Lee},
Title = {{Stablecoin Runs and the Centralization of Arbitrage}},
Journal = {Review of Financial Studies},
Volume = {Forhtcoming},
Year = {2023},
Month = {},
Pages = {},
DOI = {},
URL = {}
}

@article{Bertsch,
Author = {Bertsch, Christoph},
Title = {{Stablecoins: Adoption and Fragility}},
Journal = {SSRN Electronic Journal},
Volume = {},
Year = {2023},
Month = {},
Pages = {},
DOI = {},
URL = {https://ssrn.com/abstract=4466431}
}

@article{LiaoCara,
Author = {Liao, Gordon and Caramichael, John},
Title = {{Stablecoins: Growth Potential and Impact on Banking}},
Journal = {BGFRS International Finance Discussion Papers},
Volume = {1334},
Year = {2022},
Month = {},
Pages = {},
DOI = {},
URL = {}
}

@article{LiuEtAl,
Author = {Liu, Jiageng and Makarov, Igor and Schoar, Antoinette},
Title = {{Anatomy of a Run: The Terra
Luna Crash}},
Journal = {NBER Working Paper},
Volume = {31160},
Year = {2023},
Month = {},
Pages = {},
DOI = {},
URL = {}
}

@article{LVN,
Author = {Lyons, Richard K. and Viswanath-Natraj, Ganesh},
Title = {What keeps stablecoins stable?},
Journal = {Journal of International Money and Finance},
Volume = {131},
Year = {March 2023},
Month = {},
Pages = {},
DOI = {},
URL = {}
}

@article{CvD,
Author = {Carlsson, Hans and van Damme, Eric},
Title = {{Global Games and Equilibrium Selection}},
Journal = {Econometrica},
Volume = {61 (5)},
Year = {1993},
Month = {September},
Pages = {989--1018},
DOI = {},
URL = {}
}

@article{MS98,
Author = {Morris, Stephen and Shin, Hyun Song},
Title = {{Unique Equilibrium in a Model of Self-Fulfilling Currency Attacks}},
Journal = {American Economic Review},
Volume = {88 (3)},
Year = {1998},
Month = {June},
Pages = {},
DOI = {},
URL = {}
}

\newpage


\appendix

\section{Proofs}

\subsection{Proof of Proposition 2.1}

Suppose that there is no large sale. Observe that $\bar\theta_0$ is the unique value of $\theta$ solving
\begin{align*}
    \theta-(1-\delta)(1-F_{\bar x,\sigma_x}(\theta))=0
\end{align*}
and $x^\star_0$ is any value of $\bar x$ that solves
\begin{align*}
    F_{\mu_p(\bar x),\sigma_p}(\bar\theta_0(\bar x))-\frac{\eta}{1+\eta}=0.
\end{align*}

We first show that $F_{\mu_p(\bar x),\sigma_p}(\bar\theta_0(\bar x))$ is decreasing in $\bar x$.
By the Implicit Function Theorem,
\begin{align*}
    \frac{d\bar\theta_0}{d\bar x} = -\frac{(1-\delta)\cdot\frac{\d}{\d\bar x}F_{\bar x,\sigma_x}(\bar\theta_0)}{1+(1-\delta)\cdot\frac{\d}{\d\theta}F_{\bar x,\sigma_x}(\bar\theta_0)} = \frac{\frac{1-\delta}{\sigma_x}\cdot\phi\left(\frac{\bar\theta_0-\bar x}{\sigma_x}\right)}{1+\frac{1-\delta}{\sigma_x}\cdot\phi\left(\frac{\bar\theta_0-\bar x}{\sigma_x}\right)}>0.
\end{align*}
By the Chain Rule,
\begin{align*}
    \frac{d}{d\bar x}F_{\mu_p(\bar x),\sigma_p}(\bar\theta_0(\bar x)) &= \frac{\d}{\d\bar x}F_{\mu_p(\bar x),\sigma_p}(\bar\theta_0(\bar x))+\frac{\d}{\d\theta}F_{\mu_p(\bar x),\sigma_p}(\bar\theta_0(\bar x))\cdot\frac{d\bar\theta_0}{d\bar x} \\[0.25\baselineskip]
    &= -\frac{\mu'_p(\bar x)}{\sigma_p}\cdot\phi\left(\frac{\bar\theta_0-\mu_p(\bar x)}{\sigma_p}\right)+\frac{1}{\sigma_p}\cdot\phi\left(\frac{\bar\theta_0-\mu_p(\bar x)}{\sigma_p}\right)\cdot\frac{d\bar\theta_0}{d\bar x} \\[0.25\baselineskip]
    &= \frac{1}{\sigma_p}\cdot\phi\left(\frac{\bar\theta_0-\mu_p(\bar x)}{\sigma_p}\right)\cdot\left(\frac{d\bar\theta_0}{d\bar x}-\mu_p'(\bar x)\right),
\end{align*}
so its sign depends on $d\bar\theta_0/d\bar x-\mu_p'(\bar x)$. Observe that
\begin{align*}
    \frac{d\bar\theta_0}{d\bar x}-\mu_p'(\bar x)&=\frac{\frac{1-\delta}{\sigma_x}\cdot\phi\left(\frac{\bar\theta_0-\bar x}{\sigma_x}\right)}{1+\frac{1-\delta}{\sigma_x}\cdot\phi\left(\frac{\bar\theta_0-\bar x}{\sigma_x}\right)}-\frac{\sigma^2}{\sigma^2+\sigma_x^2} \\[0.25\baselineskip]
    &\leq \frac{\frac{1-\delta}{\sqrt{2\pi}\sigma_x}}{1+\frac{1-\delta}{\sqrt{2\pi}\sigma_x}}-\frac{\sigma^2}{\sigma^2+\sigma_x^2},
\end{align*}
since the maximum value of a standard normal density is $1/\sqrt{2\pi}$. It follows that if $(1-\delta)\sigma_x<\sqrt{2\pi}\sigma^2$, then $d\bar\theta_0/d\bar x-\mu_p'(\bar x)<0$. This holds by Assumption 1, so $F_{\bar x,\sigma_x}(\theta)$ is decreasing in $\bar{x}$. 

We check the limiting behavior of $F_{\mu_p(\bar x),\sigma_p}(\bar\theta_0(\bar x))$. Note that $1-F_{\bar x,\sigma_x}(\theta)\to0$ as $\bar x\to-\infty$, so $\bar\theta_0\to0$ as $\bar x\to-\infty$. Since $\mu_p(\bar x)\to-\infty$ as $\bar x\to-\infty$ also, we have $F_{\mu_p(\bar x),\sigma_p}(\bar\theta_0(\bar x))$ as $\bar x\to-\infty$. Conversely, note that $1-F_{\bar x,\sigma_x}(\theta)\to1$ as $\bar x\to\infty$, so $\bar\theta_0\to1-\delta$ as $\bar x\to\infty$. Since $\mu_p(\bar x)\to\infty$ as $\bar x\to\infty$ also, we have $F_{\mu_p(\bar x),\sigma_p}(\bar(\theta_0(\bar x))\to0$ as $\bar x\to\infty$. Since $F_{\mu_p(\bar x),\sigma_p}(\bar\theta_0(\bar x))$ is decreasing (and continuous) with limits $1$ and $0$ as $\bar x$ tends to $-\infty$ and $\infty$, respectively, $x^\star_0$ exists and is unique.


Suppose that a large sale occurs. Recall that $\bar\theta_1$ is the unique value of $\theta$ solving
\begin{align*}
    \theta-\delta-(1-\delta)(1-F_{\bar x,\sigma_x}(\theta))=0
\end{align*}
and $x^\star_1$ is any value of $\bar x$ that solves
\begin{align*}
    F_{\mu_p(\bar x),\sigma_p}(\bar\theta_1(\bar x))-\frac{\eta}{1+\eta}=0.
\end{align*}

We first show that $F_{\mu_p(\bar x),\sigma_p}(\bar\theta_1(\bar x))$ is decreasing in $\bar x$.
By the Implicit Function Theorem,
\begin{align*}
    \frac{d\bar\theta_1}{d\bar x} = -\frac{(1-\delta)\cdot\frac{\d}{\d\bar x}F_{\bar x,\sigma_x}(\bar\theta_1)}{1+(1-\delta)\cdot\frac{\d}{\d\theta}F_{\bar x,\sigma_x}(\bar\theta_1)} = \frac{\frac{1-\delta}{\sigma_x}\cdot\phi\left(\frac{\bar\theta_1-\bar x}{\sigma_x}\right)}{1+\frac{1-\delta}{\sigma_x}\cdot\phi\left(\frac{\bar\theta_1-\bar x}{\sigma_x}\right)}>0.
\end{align*}
By the Chain Rule,
\begin{align*}
    \frac{d}{d\bar x}F_{\mu_p(\bar x),\sigma_p}(\bar\theta_1(\bar x)) &= \frac{\d}{\d\bar x}F_{\mu_p(\bar x),\sigma_p}(\bar\theta_1(\bar x))+\frac{\d}{\d\theta}F_{\mu_p(\bar x),\sigma_p}(\bar\theta_1(\bar x))\cdot\frac{d\bar\theta_1}{d\bar x} \\[0.25\baselineskip]
    &= -\frac{\mu'_p(\bar x)}{\sigma_p}\cdot\phi\left(\frac{\bar\theta_1-\mu_p(\bar x)}{\sigma_p}\right)+\frac{1}{\sigma_p}\cdot\phi\left(\frac{\bar\theta_1-\mu_p(\bar x)}{\sigma_p}\right)\cdot\frac{d\bar\theta_1}{d\bar x} \\[0.25\baselineskip]
    &= \frac{1}{\sigma_p}\cdot\phi\left(\frac{\bar\theta_1-\mu_p(\bar x)}{\sigma_p}\right)\cdot\left(\frac{d\bar\theta_1}{d\bar x}-\mu_p'(\bar x)\right),
\end{align*}
so its sign depends on $d\bar\theta_1/d\bar x-\mu_p'(\bar x)$. Observe that
\begin{align*}
    \frac{d\bar\theta_1}{d\bar x}-\mu_p'(\bar x)&=\frac{\frac{1-\delta}{\sigma_x}\cdot\phi\left(\frac{\bar\theta_1-\bar x}{\sigma_x}\right)}{1+\frac{1-\delta}{\sigma_x}\cdot\phi\left(\frac{\bar\theta_1-\bar x}{\sigma_x}\right)}-\frac{\sigma^2}{\sigma^2+\sigma_x^2} \\[0.25\baselineskip]
    &\leq \frac{\frac{1-\delta}{\sqrt{2\pi}\sigma_x}}{1+\frac{1-\delta}{\sqrt{2\pi}\sigma_x}}-\frac{\sigma^2}{\sigma^2+\sigma_x^2},
\end{align*}
since the maximum value of a standard normal density is $1/\sqrt{2\pi}$. It follows that if $(1-\delta)\sigma_x<\sqrt{2\pi}\sigma^2$, then $d\bar\theta_1/d\bar x-\mu_p'(\bar x)<0$. This holds by Assumption 1, so $F_{\bar x,\sigma_x}(\theta)$ is decreasing in $\bar{x}$. 

We check the limiting behavior of $F_{\mu_p(\bar x),\sigma_p}(\bar\theta_1(\bar x))$. Note that $1-F_{\bar x,\sigma_x}(\theta)\to0$ as $\bar x\to-\infty$, so $\bar\theta_1\to\delta$ as $\bar x\to-\infty$. Since $\mu_p(\bar x)\to-\infty$ as $\bar x\to-\infty$ also, we have $F_{\mu_p(\bar x),\sigma_p}(\bar\theta_1(\bar x))$ as $\bar x\to-\infty$. Conversely, note that $1-F_{\bar x,\sigma_x}(\theta)\to1$ as $\bar x\to\infty$, so $\bar\theta_1\to1$ as $\bar x\to\infty$. Since $\mu_p(\bar x)\to\infty$ as $\bar x\to\infty$ also, we have $F_{\mu_p(\bar x),\sigma_p}(\bar(\theta_1(\bar x))\to0$ as $\bar x\to\infty$. Since $F_{\mu_p(\bar x),\sigma_p}(\bar\theta_1(\bar x))$ is decreasing (and continuous) with limits $1$ and $0$ as $\bar x$ tends to $-\infty$ and $\infty$, respectively, $x^\star_1$ exists and is unique.


It remains to show that $x_0^\star<x^\star_1$. Since $\bar\theta_0(\bar x)$ satisfies $\bar\theta_0(\bar x)-(1-\delta)(1-F_{\bar x,\sigma_x}(\bar\theta_0(\bar x)))=0$ by definition, we have that
\begin{align*}
    \bar\theta_0(\bar x)-\delta-(1-\delta)(1-F_{\bar x,\sigma_x}(\bar\theta_0(\bar x)))<0.
\end{align*}
Since $\theta-\delta-(1-\delta)(1-F_{\bar x,\sigma_x}(\theta))$ increases in $\theta$, it follows that $\bar\theta_0(\bar x)<\bar\theta_1(\bar x)$. Now suppose that
\begin{align*}
    F_{\mu_p(x^\star_0),\sigma_p}(\theta_0(x^\star_0))=\frac{\eta}{1+\eta}.
\end{align*}
Since $\bar\theta_0(x^\star_0)<\bar\theta_1(x^\star_0)$, 
\begin{align*}
    F_{\mu_p(x^\star_0),\sigma_p}(\theta_1(x^\star_0))>\frac{\eta}{1+\eta}.
\end{align*}
Since $F_{\mu_p(\bar x),\sigma_p}(\theta_1(\bar x))$ is decreasing in $\bar x$ under Assumption 1, it follows that $x^\star_0<x^\star_1$.

\subsection{Proof of Proposition 3.1}

Suppose that there is no large sale. By the Implicit Function Theorem,
\begin{align*}
    \frac{d\bar\theta_0}{d\delta} = -\frac{1-F_{\bar x,\sigma_x}(\bar\theta_0)}{1+(1-\delta)\cdot f_{\bar x,\sigma_x}(\bar\theta_0)}<0.
\end{align*}
By the Chain Rule,
\begin{align*}
    \frac{d}{d\delta}F_{\mu_p(\bar x),\sigma_p}(\bar\theta_0(\bar x))=\frac{\d}{\d\theta}F_{\mu_p(\bar x),\sigma_p}(\bar\theta_0(\bar x))\cdot\frac{d\theta_0}{d\delta}=f_{\mu_p(\bar x),\sigma_p}(\bar\theta_0(\bar x))\cdot\frac{d\theta_0}{d\delta}<0
\end{align*}
By the Implicit Function Theorem,
\begin{align*}
    \frac{dx^\star_0}{d\delta} = -\frac{\frac{d}{d\delta}F_{\mu_p(\bar x),\sigma_p}(\bar\theta_0(\bar x))}{\frac{d}{d\bar x}F_{\mu_p(\bar x),\sigma_p}(\bar\theta_0(\bar x))}
\end{align*}
By Assumption 1 and the proof of Proposition 2.1, the denominator is negative. Since the numerator is negative, it follows that $dx^\star_0/d\delta<0$. By the Chain Rule,
\begin{align*}
    \frac{dR_a}{d\delta} = \frac{d}{d\delta}F_{\mu,\sigma}(\bar\theta_0(x^\star_0)) = \frac{\d}{\d\theta}F_{\mu,\sigma}(\bar\theta_0(x^\star_0))\cdot\frac{d\bar\theta_0}{d\bar x}\cdot\frac{dx^\star_0}{d\delta}=f_{\mu,\sigma}(\bar\theta_0(x^\star_0))\cdot\frac{d\bar\theta_0}{d\bar x}\cdot\frac{dx^\star_0}{d\delta}<0.
\end{align*}

Suppose that a large sale occurs. By the Implicit Function Theorem,
\begin{align*}
    \frac{d\bar\theta_1}{d\delta} = \frac{F_{\bar x,\sigma_x}(\bar\theta_1)}{1+(1-\delta)\cdot f_{\bar x,\sigma_x}(\bar\theta_1)}>0.
\end{align*}
By the Chain Rule,
\begin{align*}
    \frac{d}{d\delta}F_{\mu_p(\bar x),\sigma_p}(\bar\theta_1(\bar x))=\frac{\d}{\d\theta}F_{\mu_p(\bar x),\sigma_p}(\bar\theta_1(\bar x))\cdot\frac{d\theta_1}{d\delta}=f_{\mu_p(\bar x),\sigma_p}(\bar\theta_1(\bar x))\cdot\frac{d\theta_1}{d\delta}>0
\end{align*}
By the Implicit Function Theorem,
\begin{align*}
    \frac{dx^\star_1}{d\delta} = -\frac{\frac{d}{d\delta}F_{\mu_p(\bar x),\sigma_p}(\bar\theta_1(\bar x))}{\frac{d}{d\bar x}F_{\mu_p(\bar x),\sigma_p}(\bar\theta_1(\bar x))}
\end{align*}
By Assumption 1 and the proof of Proposition 2.1, the denominator is negative. Since the numerator is positive, it follows that $dx^\star_1/d\delta>0$. By the Chain Rule,
\begin{align*}
    \frac{dR_a}{d\delta} = \frac{d}{d\delta}F_{\mu,\sigma}(\bar\theta_1(x^\star_1)) = \frac{\d}{\d\theta}F_{\mu,\sigma}(\bar\theta_1(x^\star_1))\cdot\frac{d\bar\theta_1}{d\bar x}\cdot\frac{dx^\star_1}{d\delta}=f_{\mu,\sigma}(\bar\theta_1(x^\star_1))\cdot\frac{d\bar\theta_1}{d\bar x}\cdot\frac{dx^\star_1}{d\delta}>0.
\end{align*}

\subsection{Proof of Proposition 3.2}

Fix $a\in\{0,1\}$ and hold $\tau_x$ fixed. Let $q\equiv \Phi^{-1}\!\left(\frac{\eta}{1+\eta}\right)<0$ and
let $(x_a^*(\tau),\Theta_a(\tau))$ denote the equilibrium cutoff and insolvency threshold in state $a$, where
$\Theta_a(\tau)=\bar\theta_a(x_a^*(\tau),\tau)$.

Define $z\equiv \sqrt{\tau\tau_x}\,(x-\Theta)$. The equilibrium pair $(\Theta_a,x_a^*)$ solves the system
\begin{align}
g_a(\Theta,x,\tau)
&\equiv \Theta-a\delta-(1-\delta)\Phi(z)=0,
\label{eq:g_pub}
\\
h(\Theta,x,\tau)
&\equiv \sqrt{\tau+\tau_x}\left(\Theta-\frac{\tau\mu+\tau_x x}{\tau+\tau_x}\right)-q=0.
\label{eq:h_pub}
\end{align}
Differentiating \eqref{eq:g_pub}--\eqref{eq:h_pub} with respect to $\tau$ , we have
\[
\begin{pmatrix}
g_\Theta & g_x\\
h_\Theta & h_x
\end{pmatrix}
\begin{pmatrix}
\dfrac{d\Theta_a}{d\tau}\\[4pt]
\dfrac{dx_a^*}{d\tau}
\end{pmatrix}
+
\begin{pmatrix}
g_\tau\\
h_\tau
\end{pmatrix}
=
\begin{pmatrix}
0\\0
\end{pmatrix}.
\]
By Cramer's rule,
\begin{gather}\label{eq:dTheta_pub_cramer}
\frac{d\Theta_a}{d\tau}
=
\frac{-g_\tau h_x+g_x h_\tau}{\Delta_a},
\end{gather}
where $\Delta_a\equiv g_\Theta h_x-g_x h_\Theta$, with all partial derivatives evaluated at $(\Theta_a,x_a^*,\tau)$. From \eqref{eq:g_pub} with $z=\sqrt{\tau\tau_x}(x-\Theta)$, we have
\begin{gather*}
g_x=-(1-\delta)\varphi(z)\sqrt{\tau\tau_x}, \\[0.25\baselineskip]
g_\Theta=1+(1-\delta)\varphi(z)\sqrt{\tau\tau_x}, \\[0.25\baselineskip]
g_\tau=-(1-\delta)\varphi(z)\cdot\frac{\sqrt{\tau_x}}{2\sqrt{\tau}}(x-\Theta).
\end{gather*}
From \eqref{eq:h_pub},
\begin{gather*}
   h_\Theta=\sqrt{\tau+\tau_x},\\[0.25\baselineskip]
    h_x=-\frac{\tau_x}{\sqrt{\tau+\tau_x}}. 
\end{gather*}
To compute $h_\tau$, write $\mu_p(x)=\frac{\tau\mu+\tau_x x}{\tau+\tau_x}$.
Then $\partial_\tau\mu_p(x)=\frac{\tau_x(\mu-x)}{(\tau+\tau_x)^2}$ and
\[
h_\tau=\frac{1}{2\sqrt{\tau+\tau_x}}(\Theta-\mu_p(x))-\sqrt{\tau+\tau_x}\cdot\partial_\tau\mu_p(x).
\]
Using $h(\Theta_a,x_a^*,\tau)=0$, we have $\Theta_a-\mu_p(x_a^*)=q/\sqrt{\tau+\tau_x}$, hence
\begin{equation}\label{eq:h_tau_pub}
h_\tau
=
\frac{q}{2(\tau+\tau_x)}+\frac{\tau_x(x_a^*-\mu)}{(\tau+\tau_x)^{3/2}}.
\end{equation}
Eliminating $x_a^*$ using $h=0$ yields
\[
\Theta_a=\frac{\tau\mu+\tau_x x_a^*}{\tau+\tau_x}+\frac{q}{\sqrt{\tau+\tau_x}}
\quad\Longrightarrow\quad
x_a^*
=
\frac{\tau+\tau_x}{\tau_x}\Theta_a-\frac{\tau}{\tau_x}\mu-\frac{\sqrt{\tau+\tau_x}}{\tau_x}q.
\]
Thus
\begin{gather}
x_a^*-\Theta_a
=\frac{\tau}{\tau_x}(\Theta_a-\mu)-\frac{\sqrt{\tau+\tau_x}}{\tau_x}q,
\label{eq:x_minus_Theta_pub}
\\[0.25\baselineskip]
x_a^*-\mu
=\frac{\tau+\tau_x}{\tau_x}(\Theta_a-\mu)-\frac{\sqrt{\tau+\tau_x}}{\tau_x}q.
\label{eq:x_minus_mu_pub}
\end{gather}
Substituting the above expressions into the numerator of \eqref{eq:dTheta_pub_cramer} and simplifying yields
\begin{equation}\label{eq:num_factor_pub}
-g_\tau h_x+g_x h_\tau
=
(1-\delta)\varphi(z)\,\frac{\sqrt{\tau_x}}{2\sqrt{\tau}(\tau+\tau_x)}\,
\Big(3\tau\sqrt{\tau+\tau_x}(\mu-\Theta_a)+q(2\tau+\tau_x)\Big).
\end{equation}
Thus, the sign of the numerator equals the sign of
\[
N_\tau \equiv 3\tau\sqrt{\tau+\tau_x}(\mu-\Theta_a)+q(2\tau+\tau_x).
\]

The risk component is
\[
R_a(\tau)=c_a\,\Phi\!\big(\sqrt{\tau}(\Theta_a(\tau)-\mu)\big),
\]
where $c_0=1-p$ and $c_1=p$. Differentiating, we have
\[
\frac{dR_a}{d\tau}
=
c_a\,\varphi(\sqrt{\tau}(\Theta_a-\mu))\left(
\frac{\Theta_a-\mu}{2\sqrt{\tau}}+\sqrt{\tau}\frac{d\Theta_a}{d\tau}
\right).
\]
Thus $\operatorname{sign}(dR_a/d\tau)$ equals the sign of
$\frac{\Theta_a-\mu}{2\sqrt{\tau}}+\sqrt{\tau}\frac{d\Theta_a}{d\tau}$.

Assume $\Delta_a<0$. Then \eqref{eq:dTheta_pub_cramer} and \eqref{eq:num_factor_pub} imply
$\operatorname{sign}(d\Theta_a/d\tau)=-\operatorname{sign}(N_\tau)$.

\emph{Low-fundamentals region.} If $\mu<\Theta_a$, then $(\Theta_a-\mu)/(2\sqrt{\tau})>0$.
Moreover, $\mu-\Theta_a<0$ and $q<0$ imply $N_\tau<0$, hence $d\Theta_a/d\tau>0$ and the second term is positive.
Therefore $dR_a/d\tau>0$.

\emph{High-fundamentals region.} If
\[
N_\tau>0
\quad\Longleftrightarrow\quad
\mu>\Theta_a-\frac{q(2\tau+\tau_x)}{3\tau\sqrt{\tau+\tau_x}}
=
\Theta_a+\frac{2\tau+\tau_x}{3\tau\sqrt{\tau+\tau_x}}\Phi^{-1}\!\left(\frac{1}{1+\eta}\right),
\]
then $d\Theta_a/d\tau<0$. If additionally $\mu>\Theta_a$, then $(\Theta_a-\mu)/(2\sqrt{\tau})<0$ and both terms are negative,
so $dR_a/d\tau<0$. This yields the stated sufficient condition.

For intermediate values of $\mu$, the sign is not determined by these sufficient conditions.

\paragraph{A sufficient condition for $\Delta_a<0$.}
Using $g_\Theta,g_x,h_\Theta,h_x$ as above, we obtain
\[
\Delta_a
=
\frac{-\tau_x+(1-\delta)\tau\,\varphi(z)\sqrt{\tau\tau_x}}{\sqrt{\tau+\tau_x}},
\qquad z=\sqrt{\tau\tau_x}(x_a^*-\Theta_a).
\]
Thus $\Delta_a<0$ is implied by $(1-\delta)\tau^{3/2}\varphi(z)<\sqrt{\tau_x}$, and in particular by
$\sqrt{\tau_x}>\frac{(1-\delta)\tau^{3/2}}{\sqrt{2\pi}}$ since $\varphi(z)\le 1/\sqrt{2\pi}$. This is in turn implied by Assumption 2.1.

\subsection{Proof of Proposition 3.3}

Fix $a\in\{0,1\}$ and hold $\tau$ fixed. Let $q\equiv \Phi^{-1}\!\left(\frac{\eta}{1+\eta}\right)<0$ and
let $(x_a^*(\tau_x),\Theta_a(\tau_x))$ denote the equilibrium cutoff and insolvency threshold in state $a$, where
$\Theta_a(\tau_x)=\bar\theta_a(x_a^*(\tau_x),\tau_x)$.

Define $z\equiv \sqrt{\tau\tau_x}\,(x-\Theta)$. The equilibrium pair $(\Theta_a,x_a^*)$ solves the system
\begin{align}
g_a(\Theta,x,\tau_x)
&\equiv \Theta-a\delta-(1-\delta)\Phi(z)=0,
\label{eq:g_priv}
\\
h(\Theta,x,\tau_x)
&\equiv \sqrt{\tau+\tau_x}\left(\Theta-\frac{\tau\mu+\tau_x x}{\tau+\tau_x}\right)-q=0.
\label{eq:h_priv}
\end{align}
Differentiate \eqref{eq:g_priv}--\eqref{eq:h_priv} w.r.t.\ $\tau_x$ (holding $\tau$ fixed):
\[
\begin{pmatrix}
g_\Theta & g_x\\
h_\Theta & h_x
\end{pmatrix}
\begin{pmatrix}
\dfrac{d\Theta_a}{d\tau_x}\\[4pt]
\dfrac{dx_a^*}{d\tau_x}
\end{pmatrix}
+
\begin{pmatrix}
g_{\tau_x}\\
h_{\tau_x}
\end{pmatrix}
=
\begin{pmatrix}
0\\0
\end{pmatrix}.
\]
By Cramer's rule,
\begin{equation}\label{eq:dTheta_priv_cramer}
\frac{d\Theta_a}{d\tau_x}
=
\frac{-g_{\tau_x}h_x+g_x h_{\tau_x}}{\Delta_a},
\qquad
\Delta_a\equiv g_\Theta h_x-g_x h_\Theta,
\end{equation}
with all derivatives evaluated at $(\Theta_a,x_a^*,\tau_x)$.

From \eqref{eq:g_priv} with $z=\sqrt{\tau\tau_x}(x-\Theta)$, we have
\begin{gather*}
    g_x=-(1-\delta)\varphi(z)\sqrt{\tau\tau_x}, \\[0.25\baselineskip]
g_\Theta=1+(1-\delta)\varphi(z)\sqrt{\tau\tau_x},
\\[0.25\baselineskip]
g_{\tau_x}=-(1-\delta)\varphi(z)\cdot\frac{\sqrt{\tau}}{2\sqrt{\tau_x}}(x-\Theta).
\end{gather*}
From \eqref{eq:h_priv}, we have
\begin{gather*}
    h_\Theta=\sqrt{\tau+\tau_x},\\[0.25\baselineskip]
h_x=-\frac{\tau_x}{\sqrt{\tau+\tau_x}}.
\end{gather*}
To compute $h_{\tau_x}$, write $\mu_p(x)=\frac{\tau\mu+\tau_x x}{\tau+\tau_x}$.
Then $\partial_{\tau_x}\mu_p(x)=\frac{\tau(x-\mu)}{(\tau+\tau_x)^2}$ and
\[
h_{\tau_x}=\frac{1}{2\sqrt{\tau+\tau_x}}(\Theta-\mu_p(x))-\sqrt{\tau+\tau_x}\cdot\partial_{\tau_x}\mu_p(x).
\]
Using $h(\Theta_a,x_a^*,\tau_x)=0$, we have $\Theta_a-\mu_p(x_a^*)=q/\sqrt{\tau+\tau_x}$, hence
\begin{equation}\label{eq:h_taux_priv}
h_{\tau_x}
=
\frac{q}{2(\tau+\tau_x)}-\frac{\tau(x_a^*-\mu)}{(\tau+\tau_x)^{3/2}}.
\end{equation}
Eliminating $x_a^*$ using $h=0$ yields
\[
\Theta_a=\frac{\tau\mu+\tau_x x_a^*}{\tau+\tau_x}+\frac{q}{\sqrt{\tau+\tau_x}}
\quad\Longrightarrow\quad
x_a^*
=
\frac{\tau+\tau_x}{\tau_x}\Theta_a-\frac{\tau}{\tau_x}\mu-\frac{\sqrt{\tau+\tau_x}}{\tau_x}q.
\]
Thus, we have
\begin{gather}
x_a^*-\Theta_a
=\frac{\tau}{\tau_x}(\Theta_a-\mu)-\frac{\sqrt{\tau+\tau_x}}{\tau_x}q,
\label{eq:x_minus_Theta_priv}
\\[0.25\baselineskip]
x_a^*-\mu
=\frac{\tau+\tau_x}{\tau_x}(\Theta_a-\mu)-\frac{\sqrt{\tau+\tau_x}}{\tau_x}q.
\label{eq:x_minus_mu_priv}
\end{gather}
Substituting the above expressions into the numerator of \eqref{eq:dTheta_priv_cramer} and simplifying yields
\begin{equation}\label{eq:num_factor_priv}
-g_{\tau_x}h_x+g_x h_{\tau_x}
=
-\,(1-\delta)\varphi(z)\,\frac{\tau\sqrt{\tau\tau_x}}{2\tau_x(\tau+\tau_x)}\,
\Big(\sqrt{\tau+\tau_x}(\mu-\Theta_a)+q\Big).
\end{equation}
Thus, the sign of the numerator equals the sign of
$-\big(\sqrt{\tau+\tau_x}(\mu-\Theta_a)+q\big)$.

The risk component is
\[
R_a(\tau_x)=c_a\,F_{\mu,\tau^{-1/2}}(\Theta_a(\tau_x)),
\]
where $c_0=1-p$ and $c_1=p$. Since the ex-ante distribution $F_{\mu,\tau^{-1/2}}$ does not depend on $\tau_x$,
\[
\frac{dR_a}{d\tau_x}
=
c_a\,f_{\mu,\tau^{-1/2}}(\Theta_a)\cdot\frac{d\Theta_a}{d\tau_x}.
\]
Thus $\operatorname{sign}(dR_a/d\tau_x)=\operatorname{sign}(d\Theta_a/d\tau_x)$.

Assume $\Delta_a<0$. Then \eqref{eq:dTheta_priv_cramer} and \eqref{eq:num_factor_priv} imply
\[
\operatorname{sign}\!\left(\frac{d\Theta_a}{d\tau_x}\right)
=
\operatorname{sign}\!\Big(\sqrt{\tau+\tau_x}(\mu-\Theta_a)+q\Big).
\]
\emph{Low-fundamentals region.} If $\mu<\Theta_a$, then $\sqrt{\tau+\tau_x}(\mu-\Theta_a)+q<0$ (since $q<0$), hence $dR_a/d\tau_x<0$.
\emph{High-fundamentals region.} If $\mu>\Theta_a-\dfrac{q}{\sqrt{\tau+\tau_x}}=\Theta_a+\dfrac{\Phi^{-1}(1/(1+\eta))}{\sqrt{\tau+\tau_x}}$,
then $\sqrt{\tau+\tau_x}(\mu-\Theta_a)+q>0$, hence $dR_a/d\tau_x>0$.

For $\mu$ in the intermediate interval, the sign is not determined by these sufficient conditions.

\paragraph{A sufficient condition for $\Delta_a<0$.}
As above,
\[
\Delta_a
=
\frac{-\tau_x+(1-\delta)\tau\,\varphi(z)\sqrt{\tau\tau_x}}{\sqrt{\tau+\tau_x}},
\qquad z=\sqrt{\tau\tau_x}(x_a^*-\Theta_a).
\]
Thus $\Delta_a<0$ is implied by $\sqrt{\tau_x}>\frac{(1-\delta)\tau^{3/2}}{\sqrt{2\pi}}$. This is in turn implied by Assumption 2.1.

\subsection{Proof of Proposition 3.4}

We lastly examine the comparative statics with respect to reserve fundamentals. We have
\begin{align*}
    S_L = \frac{p\cdot F_{\mu,\sigma}(\bar\theta_1(x^\star_1))}{(1-p)\cdot F_{\mu,\sigma}(\bar\theta_0(x^\star_0))+p\cdot F_{\mu,\sigma}(\bar\theta_1(x^\star_1))} = \frac{p\cdot\bar F_{\bar\theta_1(x^\star_1),\sigma}(\mu)}{(1-p)\cdot F_{\bar\theta_0(x^\star_0)}(\mu)+p\cdot\bar F_{\bar\theta_1(x^\star_1),\sigma}(\mu)}
\end{align*}
by the symmetry of the normal distribution, where $\bar{F}$ is the survival function. Then
\begin{align*}
    \frac{\d S_L}{\d\mu} = \frac{p(1-p)(f_{\bar\theta_0,\sigma}(\mu)\cdot \bar F_{(\bar\theta_1,\sigma}(\mu))-\bar F_{\bar\theta_0(\theta_0^\star),\sigma}(\mu)\cdot f_{\bar\theta_1(x^\star_1),\sigma}(\mu))}{((1-p)\cdot F_{\bar\theta_0(x^\star_0)}(\mu)+p\cdot\bar F_{\bar\theta_1(x^\star_1),\sigma}(\mu))^2}
\end{align*}
Then $\d S_L/\d\theta>0$ if and only if 
\begin{gather*}
    \frac{f_{\bar\theta_0(x^\star_0),\sigma}(\mu)}{\bar{F}_{\bar\theta_0(x^\star_0),\sigma}(\mu)} > \frac{f_{\bar\theta_1(x^\star_1),\sigma}(\mu)}{\bar{F}_{\bar\theta_1(x^\star_1),\sigma}(\mu)} \\[0.25\baselineskip]
    \frac{f_{\mu,\sigma}(\bar\theta_0(x^\star_0))}{F_{\mu,\sigma}(\bar\theta_0(x^\star_0))} > \frac{f_{\mu,\sigma}(\bar\theta_1(x^\star_1))}{F_{\mu,\sigma}(\bar\theta_1(x^\star_1))}
\end{gather*}
where the last equivalence is due to the symmetry of the normal distribution. This condition holds since the normal distribution has a decreasing reverse hazard rate and $\bar{\theta}_0(x_0^\star)<\bar{\theta}_0(x_1^\star)<\bar\theta_1(x_1^\star)$.

\end{document}